\documentclass[12pt]{iopart}

\usepackage{graphicx}
\usepackage{amstext,amssymb}
\usepackage{xspace}
\usepackage{placeins}

\newcommand{\vectorspace}[2]{\ensuremath{\mathbb{#1}^{#2}}\xspace}
\newcommand{\set}[1]{\ensuremath{\mathbf{#1}}}
\newcommand{\norm}[1]{\ensuremath{\left|\left|{#1}\right|\right|}}
\newcommand{\figtext}[1]{\footnotesize #1}

\begin{document}

\title{Revisiting the Common Neighbour Analysis and the Centrosymmetry Parameter}
\author{Peter M Larsen}
\address{Department of Physics, Technical University of Denmark, 2800 Kgs. Lyngby, Denmark}
\ead{peter.mahler.larsen@gmail.com}
\date{}

\begin{abstract}
We review two standard methods for structural classification in simulations of crystalline phases, the Common Neighbour Analysis and the Centrosymmetry Parameter. We explore the definitions and implementations of each of their common variants, and investigate their respective failure modes and classification biases. Simple modifications to both methods are proposed, which improve their robustness, interpretability, and applicability.  We denote these variants the \emph{Interval Common Neighbour Analysis}, and the \emph{Minimum-Weight Matching Centrosymmetry Parameter}.
\end{abstract}
\maketitle

\section{Introduction}
Extraction of useful material properties in a molecular dynamics (MD) simulation of condensed phases requires that we can classify crystal structures. The lowest level of this analysis is to resolve the structure at the atomic level, which is a prerequisite for determining meso- and macro-scale properties such yield strength~\cite{schiøtz1998softening, li2010nanotwins} and hardness~\cite{kallman1993silicon}, evolution in dislocation structures~\cite{stukowski2012automated, zepedaruiz2017probing} and grain structures~\cite{panzarino2015quantitative, chesser2020distinct}, and irradiation damage~\cite{marian2003modelling}.

Two structural analysis methods are particularly notable for their ubiquity: \emph{Common Neighbour Analysis} (CNA)~\cite{faken1994systematic} and the \emph{Centrosymmetry Parameter} (CSP)~\cite{kelchner1998dislocation}. These are amongst the first structural analysis methods to exploit crystal geometry, and remain the most widely used despite the development of more advanced methods based on topology~\cite{malins2013tcc, lazar2017vorotop}, geometry~\cite{ackland2006bond, larsen2016robust, martelli2018localorder}, statistical geometric (or machine-learning) descriptors~\cite{steinhardt1983boo, bartok2013soap, spellings2018machine, ceriotti2019unsupervised, adorf2020analysis}, and persistent homology~\cite{buchet2018persistent, maroulas2019persistent}.
To a large extent, these methods have endured due to their speed, simplicity, and being either parameter-free (CNA) or single-parameter (CSP) methods. Even when not used directly, they commonly serve as benchmarks for the development of new structural analysis methods.

In this paper we revisit the CNA and CSP methods. We describe their functionality and perform a comprehensive analysis of their respective failure modes, and propose some simple modifications which fix most of them. In doing so, we extend the applicability and usefulness of both methods, and raise the baseline against which new methods can be compared.

\begin{figure}[b]
\centering
\begin{minipage}{0.24\textwidth}
\centering
\includegraphics[width=1.0\textwidth]{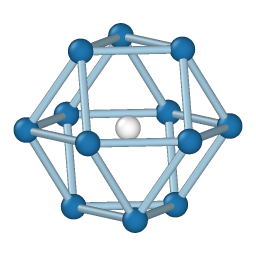}\\
{\small FCC}\\
{\small $12 \times (421)$}\\
{\small $ $}\\
\end{minipage}
\begin{minipage}{0.24\textwidth}
\centering
\includegraphics[width=1.0\textwidth]{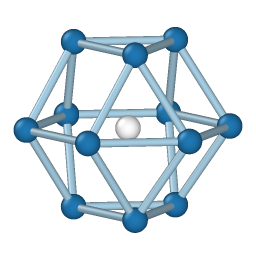}\\
{\small HCP}\\
{\small $6 \times (421)$}\\
{\small $6 \times (422)$}
\end{minipage}
\begin{minipage}{0.24\textwidth}
\centering
\includegraphics[width=1.0\textwidth]{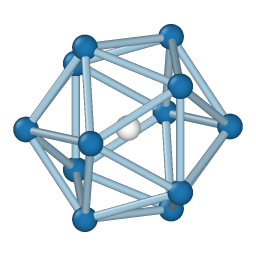}\\
{\small Icosahedral}\\
{\small $12 \times (555)$}\\
{\small $ $}\\
\end{minipage}
\begin{minipage}{0.24\textwidth}
\centering
\includegraphics[width=1.0\textwidth]{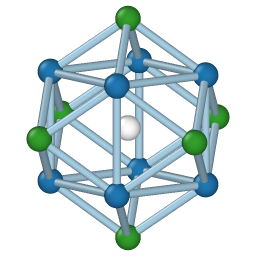}\\
{\small BCC}\\
{\small $8 \times (666)$}\\
{\small $6 \times (444)$}\\
\end{minipage}
\caption{Ball-and-stick models and CNA signatures of the local environments of the condensed phases recognized by CNA. The models consist of the central atom (white), the atoms of the first neighbour shell (blue), and (for bcc) the atoms of the second neighbour shell (green).}
\label{fig:cna_schematics}
\end{figure}

\section{Common Neighbour Analysis}
\label{sec:cna}
The CNA method has its origins in the work of Blaisten-Farojas and Andersen~\cite{blaistenbarojas1985effects}, who encoded $n$-body clusters using their common neighbour relationships.
Honeycutt and Andersen~\cite{honeycutt1987molecular} generalized this approach to a larger set of clusters and used their relative abundances to characterize structural order in nanoparticles. The currently used form of CNA is due to Faken and J{\'o}nsson~\cite{faken1994systematic}, who further extended the method to characterize the local environment of a central atom.

The CNA method employs a ball-and-stick model: each neighbour atom (ball) is joined to its nearest neighbours by a bond (stick). To identify the structure of an atom in a simulation, its ball-and-stick model is constructed and compared against the reference structures. This comparison is best described as a graph isomorphism test, but for historical reasons the comparison is made by computing a signature of common neighbour relationships. The signature consists of three indices per neighbour atom: the number of bonded neighbours shared with the central atom, the number of bonds between shared neighbours, and the number of bonds in the longest chain formed by shared neighbours.  The ball-and-stick models and their associated CNA signatures are shown in Figure~\ref{fig:cna_schematics}.

In a perfect crystal, the radial distribution function (RDF) contains well-separated peaks, which makes it easy to identify the appropriate bonded pairs. In a MD simulation, however, the atoms must necessarily move from their ideal positions. To account for the resulting variability in the distance between bonded pairs, a threshold $r_\text{c}$ is introduced which defines nearest neighbours as those whose distance satisfies $r < r_\text{c}$. Honeycutt and Andersen set $r_\text{c}$ to the first minimum in the (RDF), which lies between the first and second neighbour shells. The use of a single, global threshold is commonly referred to as the \emph{Conventional Common Neighbour Analysis} (c-CNA) method.

For some simulations it is not possible to define a single consistent threshold. For example, in multi-phase systems with differing lattice constants, the peaks in the RDF are not well-separated. To account for this variation in scale, Stukowski introduced the \emph{Adaptive Common Neighbour Analysis} (a-CNA)~\cite{stukowski2012structure}, which computes a local threshold that is specific to each atom. For reference structures containing 12 atoms, such as the fcc structure, a local length scale is calculated by averaging the distances of the atoms in the first neighbour shell:
\begin{equation}
l^\text{fcc} = \frac{1}{12}\sum\limits_{i=1}^{12} \norm{\vec{p}_i}
\end{equation}
This length scale is also employed for the hcp and icosahedral structures.
For bcc crystals, the local length scale is calculated as a weighted average of the eight atoms of the second shell and the six atoms of the third neighbour shell,
\begin{equation}
l^\text{bcc} = 
\frac{1}{14} \left( \frac{2}{\sqrt{3}} \sum\limits_{i=1}^{8} \norm{\vec{p}_i}
+ \sum\limits_{i=9}^{14} \norm{\vec{p}_i} \right)
\end{equation}
which provides an estimate of the second neighbour shell distance\footnote{This formula corrects a minor error in the original description by Stukowski.}.
The thresholds are chosen to lie, respectively, halfway between the first and second shells, and halfway between the second and third shells:
\begin{equation}
r_\text{c}^\text{fcc} = \frac{1 + \sqrt{2}}{2} l^\text{fcc},
\hspace{8mm}
r_\text{c}^\text{bcc} = \frac{1 + \sqrt{2}}{2} l^\text{bcc}
\end{equation}
Calculation of a per-atom threshold renders the a-CNA both scale-invariant and parameter-free, and is a significant improvement over the c-CNA with a global threshold. Nonetheless, the threshold is not always optimal. In fact, all existing CNA methods can fail in the following ways:
\begin{enumerate}
\item\label{failure:1} The threshold is too low, leaving some neighbour atoms under-coordinated.
\item\label{failure:2} The threshold is too high, leaving some neighbour atoms over-coordinated.
\item\label{failure:3} At a chosen threshold there are multiple signature matches, and the classification is ambiguous.
\item\label{failure:4} No consistent threshold exists: every threshold choice is simultaneously too low for some atoms and too high for other atoms.
\end{enumerate}
Here we will address points (\ref{failure:1})-(\ref{failure:3}) by considering the intervals of possible threshold choices. Point (\ref{failure:4}) is not possible to address with a CNA-type algorithm. Similarly, CNA requires that the `correct' neighbours are those which lie closest to the central atom, which is often not the case at high temperatures~\cite{larsen2016robust}. Both of these issues, though, require a more advanced structural classification approach and are not treated here.

\begin{figure}
\centering
\includegraphics[width=1.0\textwidth]{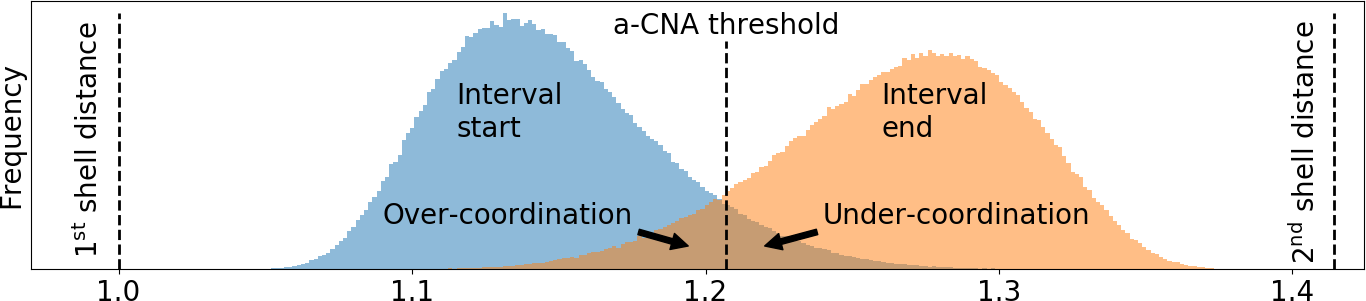}\\
$r / l^\text{fcc}$\\
\caption{Histogram of the intervals in which fcc matches can be found in a high temperature Pd simulation. The $x$-axis shows the threshold normalized by the local length scale. In a perfect fcc crystal the interval start and end points lie at $1$ and $\sqrt{2}$ respectively. The a-CNA threshold lies exactly halfway between these values. In perturbed environments the start and end points can lie on the wrong side of this threshold, resulting in miscoordinated neighbour atoms.}
\label{fig:interval_histogram}
\end{figure}

\subsection{Interval Common Neighbour Analysis}
The threshold, $r_\text{c}$, is continuous parameter. However, for any local environment, the number of threshold choices which affect the coordination environment is limited by the number of pairwise distances. 
We can investigate every meaningful threshold by inserting bonds one at a time, in sorted order (short to long). After each bond is inserted, we test for a match against a reference structure. By enumerating all thresholds choices, we map out the \emph{intervals} in which the structure is unchanged. For this reason we call this method \emph{Interval Common Neighbour Analysis} (i-CNA).

\begin{figure}
\centering
\includegraphics[width=1.0\textwidth]{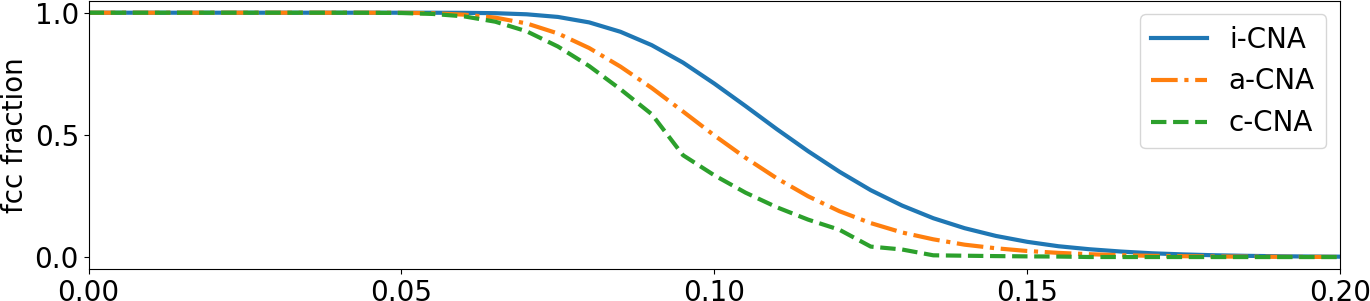}\\
$\sigma$\\
\caption{Recognition rates of the Interval CNA, Adaptive CNA, and Conventional CNA methods as a function of atomic perturbation. The atoms of a large fcc crystal with a lattice constant of $2\text{\AA}$ are perturbed by adding noise from a normal distribution $N\left( 0, \sigma \right)$. The global threshold used in the conventional CNA analysis is calculated by finding the first minimum of the RDF.}
\label{fig:fcc_fractions}
\end{figure}

\begin{figure}
\centering
{
\setlength{\fboxsep}{0pt}
\setlength{\fboxrule}{1pt}
\begin{minipage}{0.48\textwidth}
\centering
\fbox{\includegraphics[width=\textwidth]{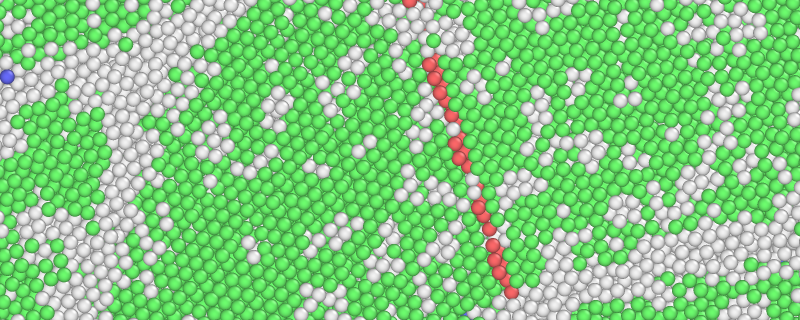}}\\
{\small a-CNA}
\end{minipage}
\hspace{1mm}
\begin{minipage}{0.48\textwidth}
\centering
\fbox{\includegraphics[width=\textwidth]{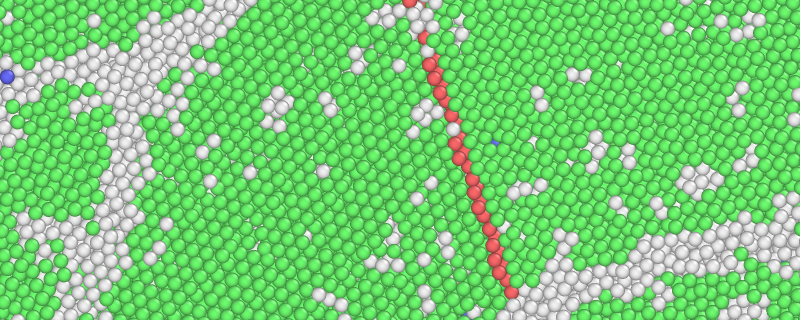}}
{\small i-CNA}
\end{minipage}
}
\caption{Comparison of Adaptive CNA and Interval CNA in a high-temperature Pd polycrystal. Interval CNA has a better recognition rate of fcc atoms ($71.4\%$ vs.~$59.1\%$, shown in green), better illustrates the twin boundary (hcp atoms, shown in red), whilst preserving the structure of the grain boundaries (disordered atoms, shown in grey). Images rendered in OVITO~\cite{stukowski2010visualization}.}
\label{fig:fcc_renders}
\end{figure}

Figure~\ref{fig:interval_histogram} shows the intervals in which fcc matches are found in a simulation of a Pd polycrystal containing 1 million atoms at $80\%$ of the melting temperature. The histograms in blue and orange show the starts and ends of the intervals. It can be seen that a significant fraction of the interval start points (end points) lie at a larger (smaller) distance than the a-CNA threshold, in which case the a-CNA threshold produces an under-coordinated (over-coordinated) structure. The i-CNA method avoids these issues by exhaustively testing the threshold intervals, which results in an optimal threshold selection for each atom.

Since the i-CNA method tests a greater number of thresholds (including an equivalent threshold to that of a-CNA), its structure recognition rate is guaranteed to be at least as good a-CNA. In order to ensure that the method does not produce false positives, we impose restrictions on the maximum threshold values:
\begin{equation}
r_\text{c}^\text{fcc} < \frac{1 + 2 \sqrt{2}}{3} l^\text{fcc},
\hspace{8mm}
r_\text{c}^\text{bcc} < \frac{1 + 2 \sqrt{2}}{3} l^\text{bcc}
\end{equation}
In the fcc (bcc) crystal, these upper bounds lie two thirds of the distance between the first and second (second and third) neighbour shells. This choice is somewhat arbitrary, but it lies at the tail end of the interval start histogram, and experimentation with a variety of simulation data reveals that a threshold which exceeds this choice tends to produce spurious classifications only.

Figure~\ref{fig:fcc_fractions} compares the recognitions rates of all three CNA methods as a function of atomic perturbation, and demonstrates that i-CNA method has a better recognition rate at all perturbation levels.  This is illustrated in MD simulation data in Figure~\ref{fig:fcc_renders}.
Due to the need to investigate multiple threshold intervals, the improved recognition comes at an additional computational cost of approximately $35\%$. The overhead is kept low by keeping track of each atom's coordination number; a full comparison is only necessary if the coordination numbers match those of a reference structure.

\begin{figure}
\centering
\includegraphics[width=1.0\textwidth]{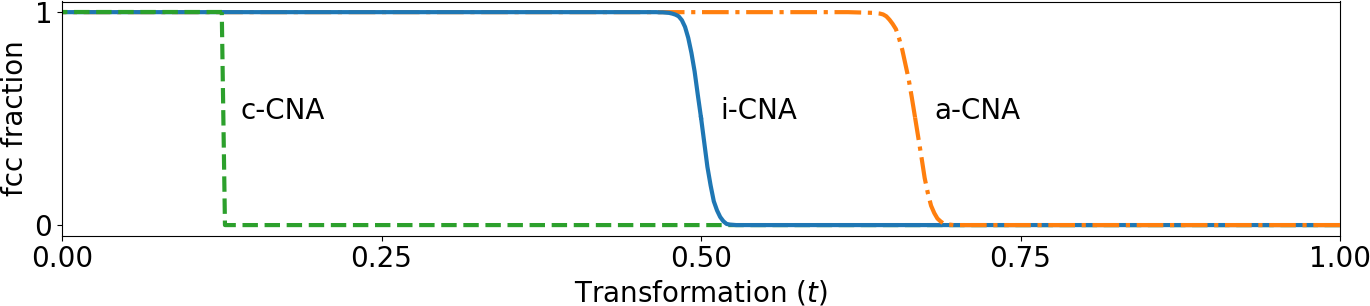}\\
\caption{Fraction of atoms identified as fcc along the Bain transformation path in a large single crystal with small atomic perturbations. The transformation is a uniaxial strain $(1-t) + t / \sqrt{2}$; the crystal is fcc at $t=0$ and bcc at $t=1$.
All atoms are either fcc or bcc (a fcc fraction of 0 means all atoms are bcc). 
The classification of the c-CNA method is biased towards bcc; the a-CNA method is biased towards fcc; the i-CNA method is unbiased.}
\label{fig:bain_fractions}
\end{figure}

\subsection{Resolving Ambiguous Classifications}
In addition to a having a good recognition rate, a structural classification method should be unbiased. This has particular importance in simulations of phase transitions, in which a biased classification can affect the conclusions drawn.

The a-CNA method tests for structure matches using two thresholds, $r_\text{c}^\text{fcc}$ and $r_\text{c}^\text{bcc}$, each of which may produce a match. The method does not have any extra information to use to resolve this ambiguity. Instead, the method (as implemented in OVITO~\cite{stukowski2010visualization}) resolves ambiguities in favour of the smaller structure. In the i-CNA method, ambiguities can be resolved by selecting the match with the widest interval.

The effect of ambiguity resolution is studied in Figure~\ref{fig:bain_fractions}. The figure shows classification rates along the Bain transformation~\cite{bain1924nature}, which is a martensitic transformation from the fcc to the bcc crystal structure that can be achieved by application of a uniaxial strain along any of the principal axes in the conventional fcc cell. The i-CNA method places the phase transition halfway along the transformation path. By favouring smaller structures, the a-CNA method places the phase transition much further along the path, and is effectively biased towards a fcc classification. In the c-CNA method, a fcc classification is not possible as soon as 14 neighbours lie within the threshold distance. Here, there is no ambiguity to resolve but the method is strongly biased towards a bcc classification.

\section{The Centrosymmetry Parameter}
\label{sec:csp}
The CSP is a structural analysis method which takes an opposite approach to CNA. Rather than using a set of reference structures to classify the topology of the local atomic environment, it computes an order parameter which quantifies the degree of inversion (or \emph{centro}) symmetry of the local environment.
In the original formulation of Kelchner~\emph{et al.}, the CSP is defined using the $N=12$ nearest neighbours of a central atom:
\begin{equation}
\text{CSP}\left(\set{R}\right) = \sum_{i=1}^{6} \norm{ \vec{r}_{i} + \vec{r}_{i+6} }^2
\label{eq:csp}
\end{equation}
where $\vec{r}_{i}$ and $\vec{r}_{i+6}$ are the vectors `corresponding to the six pairs of opposite nearest neighbours in the fcc lattice'~\cite{kelchner1998dislocation}.  For the bcc lattice the summation is replaced by the eight nearest neighbours.  For other centrosymmetric structures the appropriate summation is similarly intuitive, as each nearest neighbour atom has a clearly defined opposite neighbour.

There are two commonly used algorithms for calculating the CSP. The first algorithm (described in ref.~\cite{stukowski2012structure}) calculates a weight $w_{ij} = \norm{ \vec{r}_{i} + \vec{r}_{j} }$ for each of the $N(N-1)/2$ pairs of neighbour atoms, and calculates the CSP as the summation over the $N/2$ smallest weights. Reproduction of the Au defects described in ref.~\cite{kelchner1998dislocation} reveals that this approach was used by Kelchner \emph{et al.} 
The second algorithm (described in ref.~\cite{bulatov2006computer})
proceeds by ordering the atoms by their distance from the central atom; opposite pairs are found by (i) pairing the inner-most atom with its minimal-weight partner, (ii) removing this pair, and (iii) repeating the process until no atoms are left.

The first method is implemented in LAMMPS~\cite{plimpton1995fast} and OVITO~\cite{stukowski2010visualization}. We will describe this method as the \emph{Greedy Edge Selection} (GES) CSP. The second method is implemented in AtomEye~\cite{li2003atomeye} and Atomsk~\cite{hirel2015atomsk}. We will describe this method as the \emph{Greedy Vertex Matching} (GVM) CSP. The GVM implementation typically employs a normalization constant
\begin{equation}
\frac{1}{2 \sum\limits_i \norm{\vec{r}_i}^2}
\end{equation}
which renders the CSP invariant to scale, but, in highly centrosymmetric structures at least, these methods are otherwise equivalent.

\begin{figure}
\centering
\begin{minipage}[t]{.3\textwidth}
  \centering
  \includegraphics[width=0.85\textwidth]{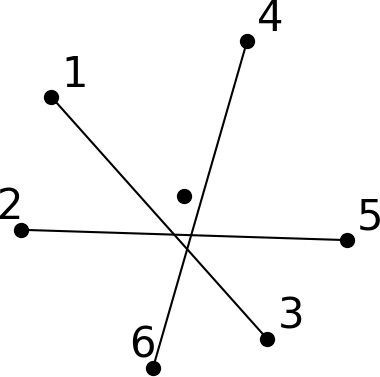}\\
  \figtext{Minimum-weight matching\\CSP=0.39}
\end{minipage}
\hspace{4mm}
\begin{minipage}[t]{.3\textwidth}
  \centering
  \includegraphics[width=0.85\textwidth]{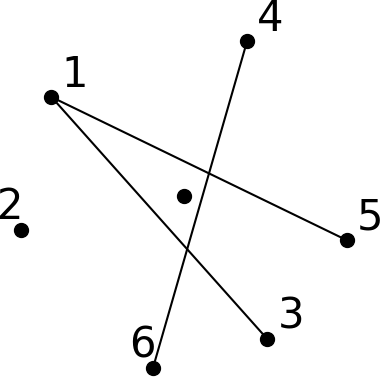}\\
  \figtext{Greedy edge selection\\CSP=0.32}
\end{minipage}
\hspace{4mm}
\begin{minipage}[t]{.3\textwidth}
  \centering
  \includegraphics[width=0.85\textwidth]{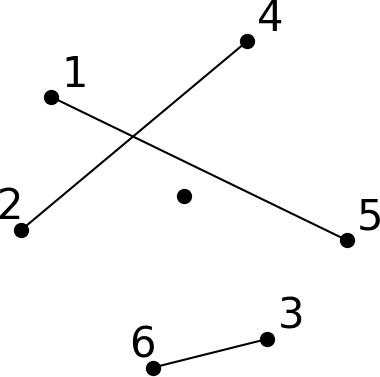}\\
  \figtext{Greedy vertex matching\\CSP=4.36}
\end{minipage}
\caption{Comparison of the MWM, GES, and GVM methods, for the nearest-neighbours of an atom in a perturbed hexagonal lattice.  The vertex labels are ordered by distance from the central atom.
}
\label{fig:csp_example}
\end{figure}

\begin{figure}
\centering
\begin{minipage}[!t]{.4\textwidth}
  \centering
  \includegraphics[width=0.6\textwidth]{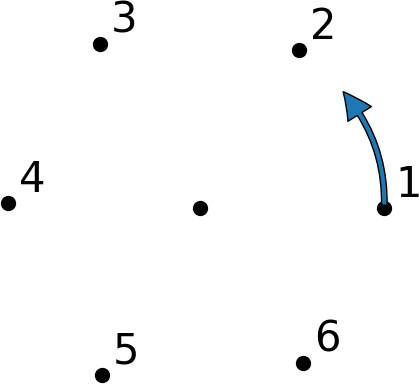}
\end{minipage}
\hspace{4mm}
\begin{minipage}[!t]{.55\textwidth}
  \centering
  \includegraphics[width=1\textwidth]{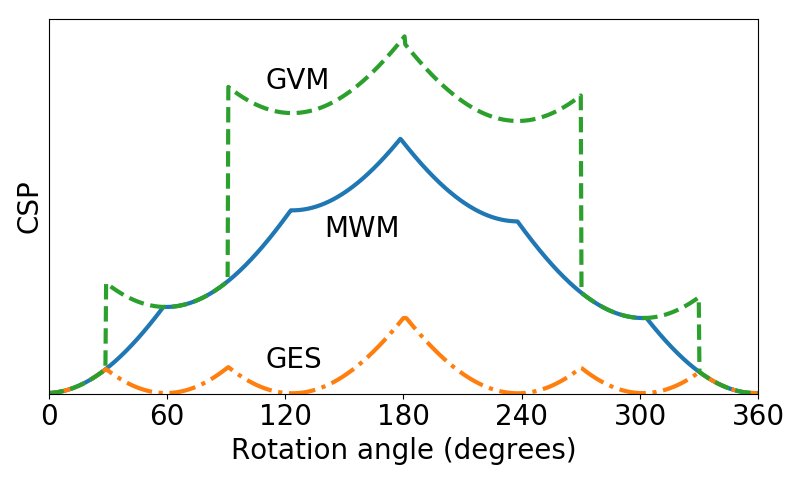}
\end{minipage}
\caption{Illustration of the failure modes of greedy CSP calculation methods.  \textbf{Left} The centrosymmetry of a slightly perturbed hexagonal structure is changed by rotating a single vertex about the central atom through an angle of $2\pi$.  Vertex labels are ordered by distance from the central atom.  \textbf{Right} The CSP values at every angle, calculated using all three methods.  GVM is not a continuous function of rotation.  GES consistently underestimates the actual CSP.}
\label{fig:csp_rotation}
\end{figure}

\subsection{The Graph Matching Centrosymmetry Parameter}
The definition in Equation~(\ref{eq:csp}) describes the CSP for centrosymmetric structures, 
but does not specify how the CSP should be calculated in the general case.
The CSP is a function $f : \vectorspace{R}{N \times d} \rightarrow \vectorspace{R}{}$ of $N$ points in $d$ dimensions, where $N$ is even. In order to generalize the CSP to arbitrary structures, we impose two conditions which must be fulfilled:
\begin{enumerate}
\item Each atom must have exactly one opposite neighbour.
\item The sum of weights must be minimal.
\end{enumerate}
If both conditions are satisfied $f$ is continuous (albeit non-differentiable) function of the input coordinates. The conditions define the CSP as a \emph{minimum-weight matching}~\cite{edmonds1965maximum} on the perfect graph with atoms as nodes and squared pairwise distances as edge weights. We will describe the calculation of the CSP satisfying the above conditions as the \emph{Minimum-Weight Matching} (MWM) CSP.

The differences in the CSP algorithms are shown for a two-dimensional example in Figure~\ref{fig:csp_example}. The MWM method produces the intuitive result. The GES method violates condition (i): some atoms have no opposite neighbours and others have multiple opposite neighbours.  The GVM method violates condition (ii): each atom has exactly one neighbour, but the weight sum is not minimal. In this example, a relatively small deviation from perfect centrosymmetry is sufficient to induce failure of the greedy methods. Furthermore, they fail in different ways and the calculated CSP values are inconsistent.

The failure modes of the greedy methods are explored in further detail in Figure~\ref{fig:csp_rotation}. The first atom is rotated through a $2\pi$ angle range, and the CSP values for each method is calculated at every rotation.  The GVM method is not a continuous function of the input coordinates; small changes in geometry cause large changes in the calculated CSP value.  The GES method, on the other hand, assigns similar CSP values for very different structures.  Calculation of the CSP as a MWM remedies both of these problems.

\begin{figure}
\centering
\includegraphics[width=1\textwidth]{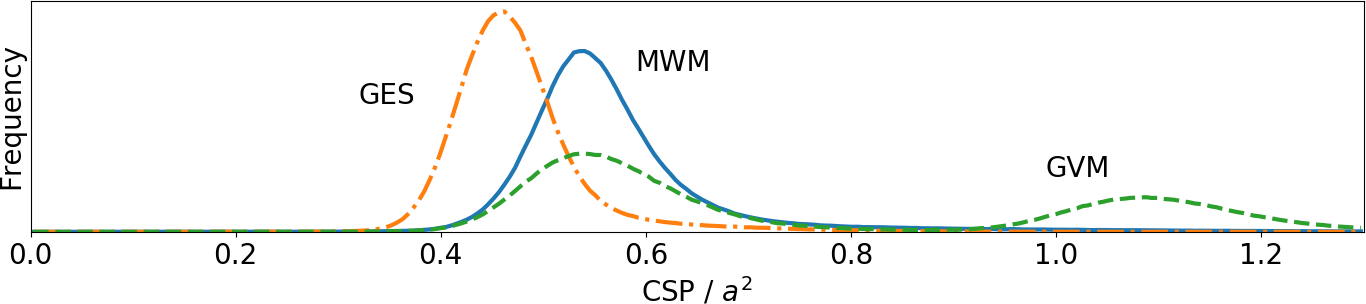}
\caption{Histograms of the CSP using the minimum-weight matching, greedy edge selection, and greedy vertex matching methods in a polycrystalline HCP Ru sample. The CSP values have been normalized by the square of the lattice constant ($a$). The GVM has a further peak at $\text{CSP} / a^2 = 2.5$, omitted here for clarity.}
\label{fig:csp_simulation}
\end{figure}

The above example is physically implausible from an energetic perspective, but the effects in a MD simulation are the same. Figure~\ref{fig:csp_simulation} shows the CSP distribution in a simulation of polycrystalline Ru in the hcp phase. The GVM method produces a distribution with multiple peaks, despite the local environments being very similar. The GES method performs better: it has a single peak, but the peak is narrower than that of the MWM method, which indicates that different local environments are mapped onto a smaller range of CSP values, and has a lower mode, which results in a poorer separation from centrosymmetric structures.

Using a standard graph matching library we achieve approximately $30,000$ CSP calculations per second on a single thread of a standard laptop computer. This is significantly slower than the greedy methods, but we can improve the running time with a hybrid strategy: first we calculate the CSP using the GES method; if the assigned graph edges constitute a valid matching then the calculated CSP is equal to the MWM method; otherwise the full MWM method is employed.  With this approach the graph matching code is only invoked in case of failure.

\section{Conclusions}
We have analyzed the behaviour of the CNA and CSP methods, and presented some simple extensions which improve their usefulness without introducing any extra parameters. By performing an exhaustive threshold search, the i-CNA method achieves a better structural recognition rate in local environments with larger atomic perturbations.  Furthermore, we introduced the Bain transformation as a test for classification bias, and it was shown that i-CNA is unbiased.

For the CSP, we demonstrated that the existing calculation methods do not produce consistent results. Using methods from graph theory, we proposed a solution which is continuous with respect to atomic perturbations and produces an intuitive result at any level of centrosymmetry.

\section*{Acknowledgments}
The author thanks for Alexander Stukowski for helpful discussions and Daniel Utt for providing Pd simulation data. This work was supported by Grant No. 7026-00126B from the Danish Council for Independent Research.

\section*{References}
\bibliographystyle{iopart-num}
\renewcommand\refname{}
\providecommand{\newblock}{}

\end{document}